\documentclass[review]{elsarticle}

\usepackage{lineno,hyperref}
\modulolinenumbers[5]

\usepackage{multibib}
\newcites{ltex}{References for Supplementary Information}

\usepackage{lineno}

\usepackage[space]{grffile}
\usepackage{eufrak}
\usepackage{amssymb,graphicx,geometry}
\usepackage{bm}
\usepackage{enumerate}
\usepackage{color}

\usepackage{times}
\usepackage{dcolumn}
\usepackage{bm}
\usepackage{amsmath}
\usepackage{mathrsfs}
\usepackage{textcomp}
\usepackage{bbold} 
\usepackage{ulem} 

\DeclareMathAlphabet{\mathpzc}{OT1}{pzc}{m}{it}

\definecolor{red}{rgb}{0.7,0,0}
\definecolor{green}{rgb}{0.,0.35,0.}
\definecolor{blue}{rgb}{0.2,0.2,0.7}
\definecolor{black}{rgb}{0.15,0.15,.15}

\newcommand{\be}{\beta^{\dagger}}

\def\be{\begin{equation}}
\def\ee{\end{equation}}
\def\bea{\begin{eqnarray}}
\def\eea{\end{eqnarray}}

\def\blue{\color{blue}}

\usepackage{lineno}
\begin{document}

\begin{frontmatter}

\title{Quantum precision measurement of two-dimensional forces with ${\bf 10^{-28}}$-Newton stability}
%\tnotetext[mytitlenote]{Fully documented templates are available in the elsarticle package on %\href{http://www.ctan.org/tex-archive/macros/latex/contrib/elsarticle}{CTAN}.}

\author{Xinxin Guo$^{{\blue \dagger}}$, Zhongcheng Yu$^{{\blue \dagger}}$, Fansu Wei, Shengjie Jin, Xuzong Chen, }
\address{State Key Laboratory of Advanced Optical Communication System and Network, School of Electronics, Peking University, Beijing 100871, China}
\author{Xiaopeng Li$^{{\blue \ast} }$}
\address{State Key Laboratory of Surface Physics, Institute of Nanoelectronics and Quantum Computing, Department of Physics, Fudan University, Shanghai 200433, China}
\address{Shanghai Qi Zhi Institute, AI Tower, Xuhui District, Shanghai 200232, China}
\address{Shanghai Research Center for Quantum Sciences, Shanghai 201315, China}
\ead{xiaopeng\underline{ }li@fudan.edu.cn}
\author{Xibo Zhang$^{{\blue \ast} }$}
\address{International Center for Quantum Materials, School of Physics, Peking University, Beijing 100871, China}
\ead{xibo@pku.edu.cn}
\author{Xiaoji Zhou$^{{\blue \ast}}$}
\address{State Key Laboratory of Advanced Optical Communication System and Network, School of Electronics, Peking University, Beijing 100871, China}
\ead{xjzhou@pku.edu.cn}




\begin{abstract}
High-precision sensing of vectorial forces has broad impact on both fundamental research and technological applications such as the examination of 
vacuum fluctuations~\cite{casimir09rmp}  and the detection of surface roughness of nanostructures~\cite{RevModPhys.89.035002}. Recent years have witnessed much progress on sensing alternating  electromagnetic forces for the rapidly advancing quantum technology---orders-of-magnitude improvement  has been accomplished on the detection sensitivity with atomic sensors~\cite{Schreppler1486,Shaniv2017,Gilmore673}, whereas precision measurement of static {electromagnetic} 
forces lags far behind with the corresponding long-term stability rarely demonstrated. Here, based on quantum atomic matter waves confined by an optical lattice, we perform precision measurement of static {electromagnetic} forces by imaging coherent wave mechanics in the reciprocal space. We achieve a state-of-the-art measurement sensitivity of  $ 2.30(8)\times 10^{-26}$~N/$\sqrt{\rm \bf Hz}$. 
%\sout{reaching a $\bf 10^7$ improvement over previous measurements of static vectorial forces in the real space~\cite{Blumseaao4453},  representing a quantum leap in the precision of force sensing.} 
Long-term stabilities on the order of $10^{-28}$~N are observed in the two spatial components of a force, which allows probing atomic Van der Waals forces at a millimeter distance~\cite{NatureNanoScanning}.  As a further illustrative application, we use our atomic sensor to calibrate the control precision of an alternating electromagnetic force applied in the experiment. 
%Unlike , which is beyond the BEC based sensing of g}
%\textcolor[rgb]{1,0,0}{The weak electromagnetic force is well controlled and the alternating force could be measured in our experiment.} 
Future developments of our method hold promise for delivering unprecedented  atom-based quantum force sensing technologies.
%\sout{ We demonstrate using our system for detecting seismic-like ground vibrations as an illustrative application.} 

\end{abstract}

%\begin{keyword}
%\texttt{elsarticle.cls}\sep \LaTeX\sep Elsevier \sep template
%\MSC[2010] 00-01\sep  99-00
%\end{keyword}

\end{frontmatter}

%\linenumbers

\newpage
Measuring force with high-precision has been attracting continuous efforts  in the interest of fundamental science discoveries and advancing technological applications~\cite{RevModPhys.89.035002}.
It plays a vital role in fundamental physics tests ranging from examination of  vacuum fluctuation mediated Casimir forces~\cite{casimir09rmp,PhysRevLett.81.4549,Chan1941} and atomic scale nanofrictions~\cite{NPGangloff} to  detecting  gravitational waves~\cite{ligo16prl,PhysRevLett.110.071105} and string theory with extra dimensions~\cite{newforce03nature}. 
It is also key to high-precision technologies such as atomic-resolution microscopy~\cite{afm03rmp,spm99rmp, mrfm95rmp,NatureMRI} and frontier nanoscale fabrication~\cite{ScienceBleszynski-Jayich272}. 
The recently upgraded SI mass unit from the macroscopic standard to the fundamental Planck constant ($\hbar$)  base~\cite{kibblebal16metrologia} creates new demands on implementation of  high-precision force measurement with quantum mechanical technologies.

As force is inherently tied to the mechanical motion in real space, high-precision force measurement has primarily relied on real space protocols, whose standard quantum limit (SQL) 
\be
 \delta F_{\rm real} \sim (m \hbar/\Delta T^3)^{1/2}, 
\label{eq:SQLreal} 
\ee 
has been established for a force $F$ acting on a free mass $m$ for a time duration $\Delta T$~\cite{1980_Zimmermann_RMP}.  
This SQL is unavoidable for mechanical motion based measurements~\cite{1975_Braginski_SQL}. 
The key to reaching high precision is to fabricate a miniaturized sensor with extremely small mass and ultralow defect density, and at the same time maintain a high level of signal-to-noise ratio~\cite{RevModPhys.89.035002}. In recent years, there has been much research effort devoted to  optimizing nanoscale opto-mechanical devices~\cite{APL2001, Teufel2009, Moser2013, moser14natnano, Rossi2017, Arcizet2017natnano,  PhysRevLett.121.063602} and quantum atomic gases~\cite{ Schreppler1486, Shaniv2017, Gilmore673, NatureNanoScanning, Biercuk2010, Blumseaao4453}  for high precision force sensing. With nanoscale opto-mechanical devices in a cryogenic environment, a force measurement sensitivity of a few times $10^{-21}$N/$\sqrt{\mathrm{Hz}}$ has been achieved~\cite{Moser2013,moser14natnano}. The atomic gas system provides an approach of quantum force sensing using a large number of microscopically identical force sensors, namely the atoms.  With atomic sensors,  a measurement sensitivity of about $3.5\times10^{-26}$~N/$\sqrt{\mathrm{Hz}}$ has been reached for alternating (a.c.) forces~\cite{Schreppler1486,Gilmore673}. \textcolor[rgb]{0,0,0}{However, for static (d.c.) electromagnetic forces, the so-far achieved sensitivity record is $10^{-19}$~N/$\sqrt{\mathrm{Hz}}$, which is obtained using  a three-dimensional trapped ion sensor~\cite{Blumseaao4453}, substantially lagging behind the a.c. force measurement.}
For an atomic Bose-Einstein condensate (BEC), an external force applied to the system produces a macroscopic quantum state of matter wave, developing  a wavevector-to-force response apart from the spatial motion. This quantum phenomenon  has been observed for alkali metal atomic matter waves with persisting quantum coherence~\cite{HCN08prl,Aidelsburger13prl,Ketterle13prl}. 
The BEC confined in a one-dimensional lattice  has been applied to measure the earth gravity along one fixed dimension~\cite{prl06Tino, PhysRevLett.106.038501}. 
 Here, we perform precise sensing of a two-dimensional force by separating the wavevector-to-force response from the real-space dynamics using a two-dimensional optical lattice, such that the spatial motion is suppressed to minimize potential systematic errors without compromising the detection signal-to-noise ratio. 
This sets our force measurement  free from the type of SQL in Eq.~\eqref{eq:SQLreal}, and we thereby accomplish a state-of-the-art sensitivity of $2.30(8)\times 10^{-26}$~N/$\sqrt{\mathrm{Hz}}$ in the measurement of two-dimensional d.c. electromagnetic forces.
With such high measurement sensitivity, 
long-term stabilities of $1.7^{+1.3}_{-0.4}\times 10^{-28}$~N and $6.3^{+1.2}_{-0.7} \times 10^{-28}$~N for the two spatial components of a vectorial force are demonstrated based on consecutive experimental runs for $3\times10^5$ seconds.  
%
%This allows detecting atomic Van der-Waals forces at a millimeter distance away~\cite{NatureNanoScanning}. 
%
%\sout{We observe the force measurements are sensitive to anomalous seismic-like vibrations around the lab, which illustrates an application of high-precision force-based quantum technology. }
\textcolor[rgb]{0,0,0}{The measurement precision for weak electromagnetic forces is sufficiently high such that an alternating force can be well controlled and measured in our experiment.}
%Our high sensitivity of measurements enables detecting anomalous seismic-like vibrations around the lab. 
%Our results lay the experimental foundation for measuring and engineering vectorial forces in two and three dimensions with $10^{-30}$~N or higher precision in the future. 
Our results represent a significant step forward in the precision of force sensing, potentially promoting the force-based fundamental physics discoveries and technological applications, including testing vacuum fluctuations~\cite{casimir09rmp,PhysRevLett.81.4549,Chan1941} and performing remote surface tomography by probing atomic Van der Waals forces~\cite{NatureNanoScanning}.

\section{Decoupling the real- and reciprocal-space dynamics with a light-crystal matter wave}

\begin{figure*}
 		\centerline{\includegraphics[angle=0,width=.9\textwidth]{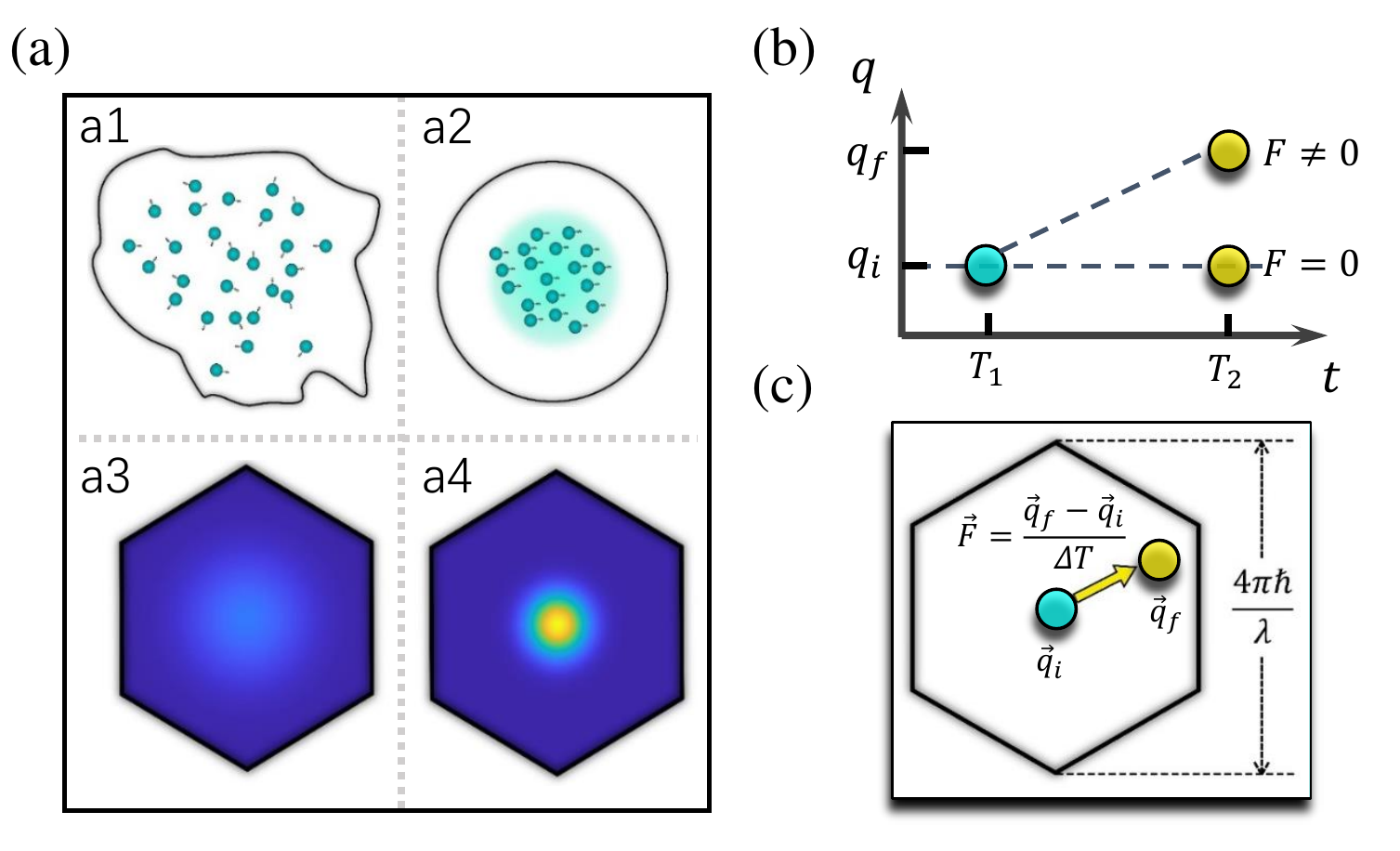}}
 	%\end{figure*}
%	\clearpage 
 %	\begin{figure*}
 		\caption{
 			{\bf High-precision force measurement by atomic BEC.}
 			{\bf (a)}, Illustration of how quantum coherence enhances sensitivity. More thermal atoms (a1) cause a momentum distribution broadening (a3) and add thermal noise to force sensing; whereas more atoms in a BEC (a2) improve the signal-to-noise ratio due to quantum coherence (a4). {\bf (b)}, Wavevector accumulation of the quantum atomic matter wave under an external force.  ({\bf c}), The reciprocal space of a triangular light crystal formed by laser. The size and the hexagon shaped boundary of the first  Brillouin zone, as determined by the laser wavelength ($\lambda$) and orientation geometry, define a rigid two-dimensional coordinate frame for precise measurement of atomic wavevector.  
 		}
 		\label{fig:f1}
 	\end{figure*}

Our experiment is based on an atomic BEC confined by a triangular two-dimensional optical lattice (Fig.~\ref{fig:f1}). 
We take each  atom as an individual force sensor and implement the impulse-momentum theorem 
\be
\vec{F} = \frac{\vec{q}_f -\vec{q}_i} {\Delta T} 
\label{eq:force} 
 \ee 
 for quantum force sensing. Here $\vec{q}_i$ ($\vec{q}_f$) is the initial (final) quasi-momentum of each atom in the light crystal, and $\Delta T$ is the time duration of the force acting on the BEC.  
As $\Delta T$ can be accurately controlled in the experiment, high precision force sensing is then reduced to detecting quasi-momentum.  
Having a BEC in  equilibrium to start with, the momentum is zero initially, and remains vanishing in the absence of an external force, which corresponds to a background-free force measurement (Fig.~\ref{fig:f1}~(b)).  
Because of the wave mechanical nature of a BEC, the quasi-momentum response under an external force can be detected by measuring the wavevector $\vec{k}_f$ of the matter wave ($\vec{k}_f = \vec{q}_f/\hbar$), which, in sharp contrast to conventional mechanical schemes~\cite{RevModPhys.89.035002,Schreppler1486,Gilmore673, 1980_Zimmermann_RMP, Blumseaao4453}, no longer requires probing real-space motion. 
%With an external force applied to the BEC, an atomic matter wave is generated with an accumulated wavevector, $\vec{q}_f/\hbar$. 
The wavevector is precisely detectable in the experiment because the atoms in the matter wave form a macroscopic quantum state  sharing the same quasi-momentum, giving rise to a sharp Bragg peak in the reciprocal space (Fig.~\ref{fig:f1}~(a4)).  

One novel aspect with our light crystal BEC sensing is the decoupling of reciprocal-space dynamics from the real-space mechanical motion. The precision force sensing solely relies on the differential wavevector measurements and the fundamental Planck constant. 
The wavevector measurement can be precisely calibrated according to the first Brillouin zone (FBZ), a natural reference frame of the reciprocal space  provided by the rigid light crystal (Fig.~\ref{fig:f1}~(c)). 
With our BEC force sensor, the fundamental limitation on the force measurement precision is from the quantum shot noise in the atomic quasi-momentum caused by the shallow harmonic trap,  
\be
\delta F_{\rm reciprocal} \sim \frac{\hbar} {\sqrt{N_0} \Delta T \times l_Q } \textrm{         }, 
\label{eq:dFrec} 
\ee  
with $l_Q$ a characteristic length scale of the BEC, and $N_0$ the number of condensed atoms~\cite{SI21Guo}. 
The measurement scheme is  free of such  demanding experimental techniques  as superresolution imaging or velocimetry as in real-space mechanical motion techniques.  Our $\delta F_{\rm reciprocal}$ is completely independent of atomic mass, spatial position and velocity, or their technical fluctuations. 
This scheme then avoids the constraint by the well celebrated SQL for real-space motion based protocols~\cite{1980_Zimmermann_RMP,RevModPhys.89.035002,Blumseaao4453,Schreppler1486,Gilmore673}. Moreover,  with the atoms strongly confined  by the light crystal, the real-space atomic motion is strongly suppressed. For the force sensing, such strong suppression minimizes the systematic errors caused by initial position drift in measuring the atomic quasi-momentum after a time-of-flight process~\cite{SI21Guo}. The resultant force sensitivity as measured in our experiment reaches $2.30(8)\times 10^{-26}$~N/$\sqrt{\mathrm{Hz}}$, with both the measurement time and the BEC preparation time included in the definition of the measurement sensitivity. Furthermore, when considering the force acting time of $\Delta T \approx 4$~ms only, we obtain a force measurement sensitivity of $1.7\times 10^{-28}$~N/$\sqrt{\mathrm{Hz}}$, 
%{\color{red}  
reaching significantly beyond the SQL for the mechanical force measurement~\cite{1980_Zimmermann_RMP} of $9\times 10^{-28}$~N/$\sqrt{\mathrm{Hz}}$ (using Eq.~\eqref{eq:SQLreal}) under the same measurement time. 
%} 

\section{Reaching long-term high stability}

In the experiment we have a laser-produced triangular light crystal that spans across the $x$-$y$ plane (Fig.~\ref{fig:f2}(a)), with the $y$-direction (vertical) along the earth gravity~\cite{Jin_2019,PhysRevLett.126.035301}.  
The atomic gas is thus confined in a two dimensional array of tubes elongated along the $z$-direction. 
%{\color{red} 
The atomic density in the tube is tunable. This allows to maintain sufficiently weak interaction, which would otherwise cause decoherence in the wave dynamics and a consequent limitation on the measurement time. 
%}
Our BEC contains about $2\times 10^5$ atoms. 
We apply external forces in the $x$-$y$ plane after preparation of the BEC system.  To demonstrate the generality of our BEC force sensing, we implement both optical and magnetic forces~\cite{SI21Guo}, to be referred to as $\vec{F}_{\rm o}$ and $\vec{F}_{\rm m}$, respectively. After the force is fully applied, we measure the wavevector of the generated matter wave by time-of-flight imaging~\cite{SI21Guo}. 
%We start to perform measurements after the force stabilizes. This avoids the difficult task of calibrating the complex out-of-equilibrium dynamics of the atomic system that happens when the forces turns on~\cite{SI21Guo}.  
By introducing feedback control of laser intensity and magnetic coil currents to afford a large number of measurement cycles, the forces are maintained stable for consecutive experimental runs over a long-term period of several days. We also make efforts on holding the stability of  experimental conditions such as room temperature and humidity to our best capability. 
Based on the experiment designing principle in Eq.~\eqref{eq:force},  our measurements are immune to the fluctuations of  atom number, lattice depth, and optical phases, some of which can be difficult to control precisely. 

By varying the force acting time $\Delta T$, we measure how the wavevector of the matter wave accumulates in its dynamics. The results are shown in Fig.~\ref{fig:f2}(b). The observed linear relationship confirms that the experimental noise is well under control such that the momentum impulse theorem (Eq.~\eqref{eq:force}) indeed applies to the BEC wave dynamics. The vectorial forces---both of the magnitude and the direction---can then be reliably extracted either by fitting the realtime wavevector accumulation or by calculating the difference between the initial and final points. We proceed by performing a differential wavevector measurement according to the second approach that only requires two experimental cycles taking about $\tau_0 \approx 76$~s.  

We determine the long-term stability of the force measurement by collecting a large amount of experimental data. Figure~\ref{fig:f2}(c) and (d) show two groups of measurements corresponding to optical and magnetic forces applied along a direction slightly deviated from the $y$ axis.  
By analyzing the results in Fig.~\ref{fig:f2}(c1, d1), we find the  data points collected over a long period of time---several days for the optical force---obey a standard normal distribution (Fig.~\ref{fig:f2}(c2, c3, d2, d3)). This implies there is no serious long-term drift in our experiment, because otherwise the distribution can develop a structure of multiple peaks or a flattened plateau in the middle. More systematically, 
the long-term stability is characterized by  the Allan Deviation (ADEV) as a function of data averaging time $\tau_1$~\cite{RevModPhys.89.035002} as shown in  Fig.~\ref{fig:f2}(c4) and (d4). 
It is evident from the ADEV of $x$-component of the optical force $F_{{\rm o},x}$ that  our force sensing remains stable up to  tens of thousands of  seconds, where the consecutive measurement fluctuations are dominated by white noise. With the high sensitivity by our BEC force sensor, we reach a long-term stability $\delta F_{{\rm o}, x} = 1.7^{+1.3}_{-0.4}\times 10^{-28} $~N (Fig.~\ref{fig:f2}(c4)).  
For the $y$-component, the ADEV has a slightly different behavior. 
It shows an upward bending behavior after $\tau_1$ exceeds a certain characteristic time. %, which reflects the time scale of long-term instability. 
%By comparison, the ADEV of the $y$-component of the optical force shows an upward bending behavior after $\tau_1$ exceeds $10^4$~s.
%
%The upward bending behavior of the ADEV observed for the $y$-component forces reflects a characteristic time scale of long-term instability.  
%
Owing to the apparent distinction between the two vectorial components, we attribute the instability in the $y$-component measurement, at a time scale of $10^4$ and  $10^3$ seconds respectively for optical and magnetic forces, to the force control rather than being intrinsic to the BEC force sensing.  
The two components of the optical force are determined as $F_{{\rm o}, x} = 1.0(2)\times 10^{-27}$~N and $F_{{\rm o} ,y} = 7.81(6)\times 10^{-26}$~N, and the direction angle of the force with respect to the $y$-axis is determined as $\theta_{\rm o} =  (0.013 \pm 0.002)$~radian. 
For the magnetic force, we measure $F_{m, x} = -2(2)\times 10^{-28}$~N, and $F_{m, y} = (5.336 \pm 0.016)\times 10^{-25}$~N, $\theta_{\rm m} =-4(4)\times 10^{-4}$~radian, correspondingly. As a crosscheck, we also show the force stabilities by the technique of binning the experimental data and analyzing the corresponding histogram~\cite{ludlow08science}, and the results are comparable to the ADEV analysis~\cite{SI21Guo}. 

\section{Sensitivity of the BEC force sensor}

We extract  the measurement sensitivity from the ADEV by taking  the ADEV averaging time $\tau_1 = \tau_0$~\cite{SI21Guo}. For the optical force with $\Delta T = 4.2$~ms, we determine a sensitivity for the $x$ and $y$ components,  $S_{\mathrm{o},x} =2.56(3)\times 10^{-26}$ N/$\sqrt{\mathrm{Hz}}$, and $S_{\mathrm{o},y} = 4.20(5)\times 10^{-26}$~N/$\sqrt{\mathrm{Hz}} $, with both the measurement and BEC preparation times included. 
For the magnetic force whose magnitude is relatively larger, we choose $\Delta T = 3.6$~ms, and measure a sensitivity $S_{\mathrm{m},x} =2.30(8)\times 10^{-26}$~N/$\sqrt{\mathrm{Hz}} $, and $S_{\mathrm{m},y} =6.1(2)\times 10^{-26}$~N/$\sqrt{\mathrm{Hz}} $.
 The BEC sensor thus demonstrates a high sensitivity on the $10^{-26}$~N/$\sqrt{\rm Hz}$ level for measuring two-dimensional electromagnetic forces, which shows significant improvement over previous d.c. measurements of vectorial electromagnetic forces~\cite{PhysRevLett.121.063602,Blumseaao4453}.

\begin{figure}
 	\centerline{\includegraphics[angle=0,width= 0.6\textwidth]{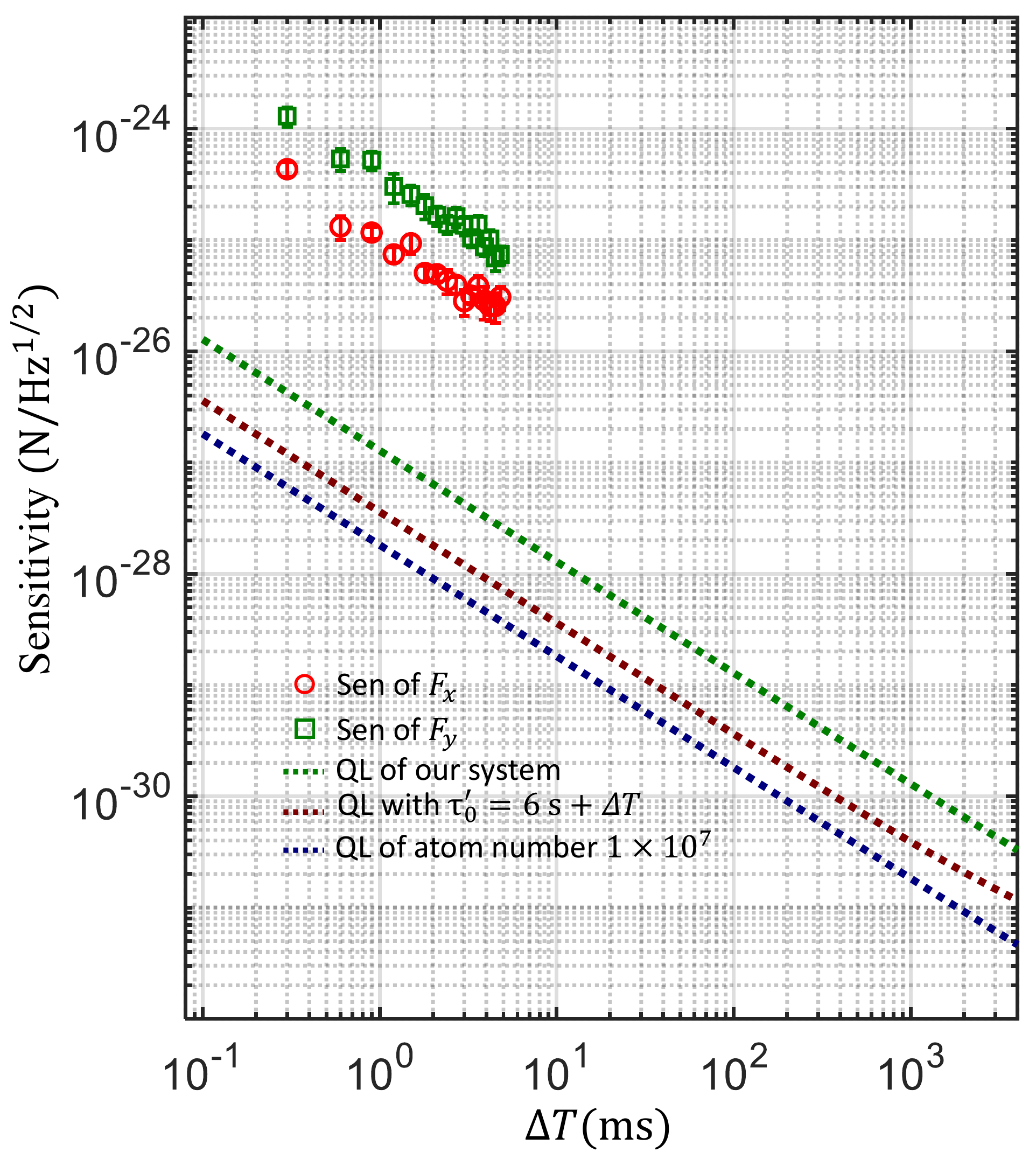}}
 %\end{figure}
%\clearpage
 %\begin{figure}
 	\caption{
 		{\bf The sensitivity of force measurement under varying experimental conditions.} The red circles and green squares correspond to the experimentally determined sensitivities for $x$- and $y$-components of the optical force with various  force acting times ($\Delta T$). Error bars represent $1\sigma$ statistical uncertainties. The green dotted line is the theoretical quantum limit (QL) of the sensitivity calculated by taking the  implemented experimental condition, which is proportional to $1/\Delta T$ as the BEC preparation time dominates over the force acting time in our experiment. The blue and brown dotted lines indicate quantum limits for an increased atom number ($10^7$) and for a reduced BEC preparation time of ($3$~s), respectively. 
		%The arrow indicates that the measurement time can be enhanced to the five-second level when the interaction-induced dephasing is significantly suppressed. 
 	}
 	\label{fig:f3}
 \end{figure}

From Eq.~\eqref{eq:dFrec}, the force sensitivity is expected to improve further upon increasing measurement time as $1/\Delta T$, in the regime where the BEC preparation time dominates over the measurement time. This scaling is confirmed in our experiment (Fig.~\ref{fig:f3}) for both spatial components of a force. It is natural to improve the force sensitivity by increasing the measurement time or reducing the BEC preparation time. The preparation time is about  $38$~s in our experiment, which can be  readily reduced to a few seconds with advanced cold atom technology, for example by dynamically reducing the trap size~\cite{PhysRevA.71.011602}. It is less trivial to increase the measurement time because this is limited by the interaction induced decoherence. That would require using a different atomic species whose interactions are negligible or can be tuned by Feshbach resonances.  Having the interaction shut off, for example with $^{133}$Cs or $^{88}$Sr atoms, we expect the measurement time can reach the order of ten seconds~\cite{HCN08prl}, and the sensitivity would then be improved by another three orders-of-magnitude. Once the interaction is off, the sensitivity can also be improved further by increasing the atom number, for which the improvement scales as $\sqrt{N_0}$.  
Although our measured sensitivity has exceeded the SQL established for mechanical schemes, there is still quite some room for us to improve the technicality. Fig.~\ref{fig:f3} shows the computed quantum limit (QL) of our quantum wave mechanical measurement, which is about a factor of 80 smaller than  the measured sensitivity of $F_{\mathrm{o},x}$. This can be attributed to residual experimental imperfections such as fluctuations in the trapping potential and environmental temperatures and humidity. Taking all these potential improvements into account, realizing a measurement beyond a  $10^{-30}$~N stability is anticipated in the near term with the BEC force sensing scheme.

%\section{ Illustrative application by detecting seismic-like ground vibrations.}

%\sout{ BEC force sensing provides novel opportunities for atomic quantum sensors. 
%The realized high measurement stability allows probing atomic Van der Waals forces at a millimeter distance~\cite{NatureNanoScanning}, which would enable unprecedented atom force based quantum sensing technology. 
%As an illustrative application, we demonstrate how to  use the BEC force sensing for detecting seismic-like ground vibrations. 
%We simulate a mini-earthquake in the lab by putting a vibrating motor on the floor. Ground vibrations
%generated by the motor are measured independently and have a peak noise power spectral density of   $2\times10^{-4}g$/$\sqrt{\mathrm{Hz}}$ ~\cite{SI21Guo}. The ground vibrations are coupled to the experimental platform and then picked up by the atomic BEC force sensor. 
%The realtime experimental data are shown in Fig.~\ref{fig:f4}. 
%In absence of the ``mini-earthquake", we measure a force of $1.6\times 10^{-25}$~N  along the vertical direction having a variance $(2.92 \pm 0.16)\times 10^{-53}$~N${}^2$. In sharp contrast, when the ``mini-earthquake" happens, the force variance becomes $8\times 10^{-52}$~N${}^2$ instantaneously, more than ten time larger than the absence of the ``min-earthquake" (Fig.~\ref{fig:f4}(b,c)).  This demonstration shows the BEC force sensing can be readily implemented for advancing  atom force based quantum technology for its high sensitivity. }

\section{Calibrating alternating electromagnetic forces by BEC sensor}  
\begin{figure}
 	\centerline{\includegraphics[angle=0,width=.9\textwidth]{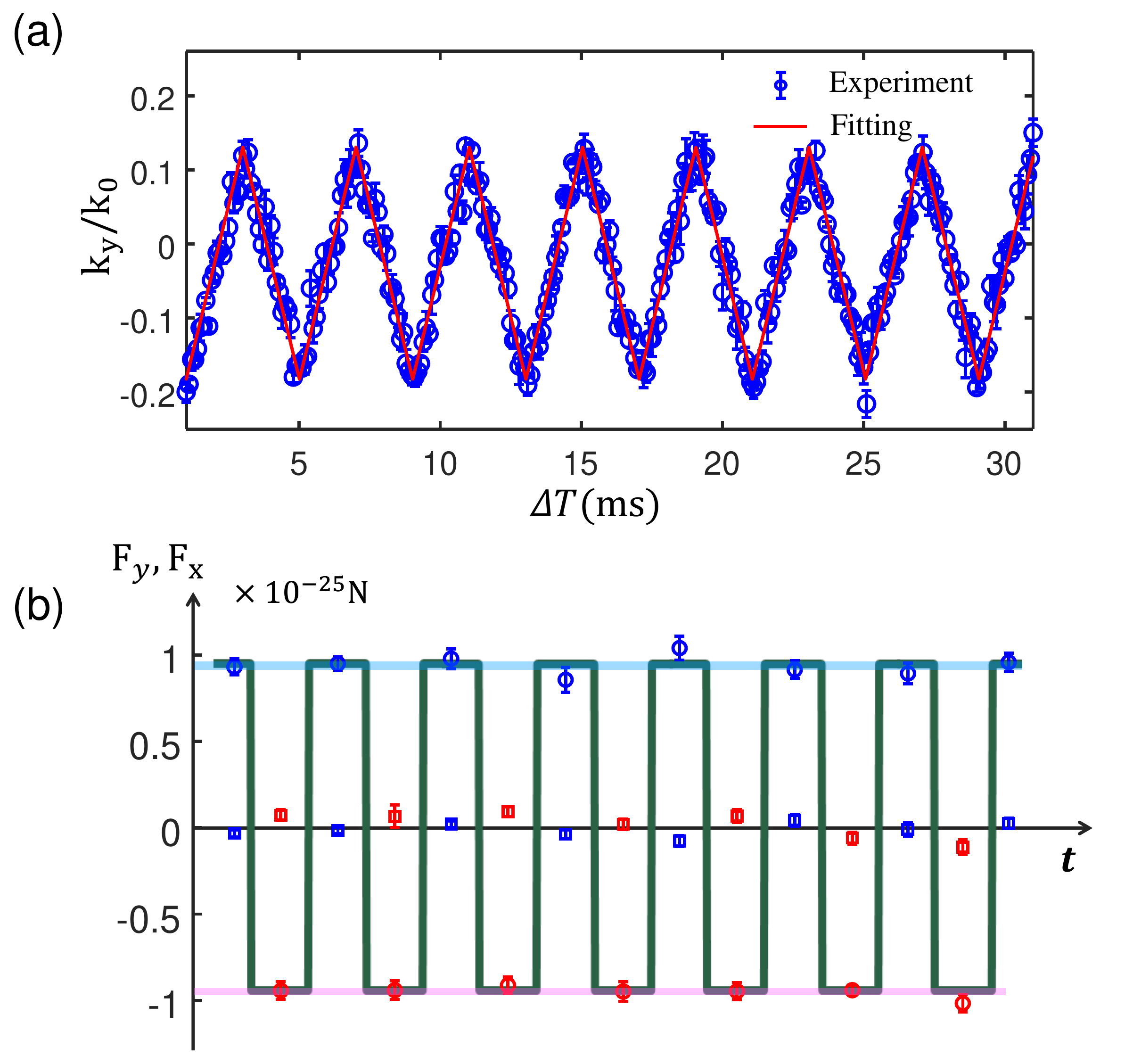}}
 	\caption{\textcolor[rgb]{0,0,0}{ {\bf BEC sensing of alternating electromagnetic forces in two dimensions.} 
	{\bf (a)},  Measurement of atomic wave evolution upon alternating forces along the $y$-direction. Blue spheres represent the experiment measurements data, and the error bars represent the standard error averaging over five independent experiment runs. Red solid lines represent the fitting of the experimental data. 
 	{\bf (b)}, Experimental results on measuring the alternating electromagnetic force. The Blue spheres (red spheres) represent the measurements of $y$-components of the alternating force. Blue squares (red squares) represent the measurements of $x$-components of the alternating force. Dark green line is a guide-to-the-eye that illustrates the periodicity for modulating the $y$-component of the force.
		}}
 	\label{fig:f4}
 \end{figure}

%High precision measurement of alternating electromagnetic forces.}
BEC force sensing provides novel opportunities for atomic quantum sensors. 
The realized high measurement stability allows probing atomic Van der Waals forces at a millimeter distance~\cite{NatureNanoScanning}, which would enable unprecedented atom force based quantum sensing technology.  As an illustrative application, we use the BEC sensor to calibrate the control precision of the electromagnetic forces applied to the atomic systems in the experiment. By periodically changing the intensity of the external optical dipole trap (Fig.~\ref{fig:f4}), we deliberately apply a square-wave modulation in the $y$-component of the force on the BEC. 
Atoms in the BEC sensor exhibit an oscillating wavevector of a triangular wave form. 
Fig.~\ref{fig:f4}(a) gives the accumulation of the $y$-component wavevector of the matter wave with different force acting time $\Delta T$. The oscillating frequency of the wavevector is determined to be $250$~Hz. The applied forces at each plateau of the square wave are extracted by fitting the experimental results of the BEC wave vector.  
The mean value of the forward direction of the force (marked by the light blue shading in Fig.~\ref{fig:f4}(b)) is $ (9.41 \pm 0.19)\times 10^{-26}$~N. 
The mean value of the force along the opposite direction (the light red shading) is $(-9.47 \pm 0.17)\times 10^{-26}$~N. 
The measured forces along the $x$-direction are one-order-of magnitude smaller.
The force variance across different periods of the square wave is 1 to 3$\times 10^{-53}$~N$^2$. These measurements confirm that our control precision for the electromagnetic forces is at the level of $10^{-27}$~N even in the a.c. regime.

\section{Conclusion}
We develop a light crystal BEC system for vectorial force sensing using quantum wave dynamics, and perform high precision measurement of two-dimensional static electromagnetic forces.  Our determined force sensitivity reaches beyond the SQL well established for real-space mechanical response.
%, \sout{and has  a $10^7$ practical improvement over the previous d.c. measurement on vectorial forces in electromagnetic or opto-mechanical settings  ~\cite{PhysRevLett.121.063602,Blumseaao4453}. }
We further achieve a long-term stability at the level of $10^{-28}$~N for measuring and controlling the force, which enables probing distant Van der Waals forces. \textcolor[rgb]{0,0,0}{The electromagnetic force can be well controlled in our system, enabling us to calibrate alternating forces.} The BEC force sensing  opens up fascinating opportunities not only for advancing fundamental science in the aspect of probing vacuum fluctuations, but also for developing atomic quantum technology such as  atom-force based imaging for nanostructures. \\
\\
\noindent\textbf{Acknowledgements}\\
This work is supported by National Program on Key Basic Research Project of China (Grant Nos. 2018YFA0305601, 2021YFA0718300, 2021YFA1400900), National Natural Science Foundation of China (Grant Nos. 61727819, 11934002, 11874073), Shanghai Municipal Science and Technology Major Project (Grant No. 2019SHZDZCX01), and the Chinese Academy of Sciences Priority Research Program (Grant No. XDB35020100). %and the Hefei National Laboratory.
\\

\noindent\textbf{Author Contributions}\\
X.P.L., X.Z., and X.J.Z. conceived  the idea of this project and initiated the experiment. 
X.G., Z.Y., F.W., S.J. constructed the BEC sensor and performed the force measurements. 
%X.B.Z. analyzed the sensitivity and the long-term stability, with X.G. and X.Y. 
All contributed to the data analysis and the writing of this manuscript. \\
%\textcolor[rgb]{1,0,0}
{${\blue \dagger}$These authors contributed equally to this work.}

%All contributed to the project and the writing of this paper. 
%X.G., Z.Y., F.W., S.J., X.C.,X.L.,X.Z. and X.J.Z. conceived, designed and carried out the experiments mentioned in this manuscript. All authors discussed the results and contributed to the writing of this manuscript.

\noindent\textbf{Author Information}\\
The authors declare no competing financial interests.

\bibliography{mybibfile}

\pagebreak

%\begin{singlespace} 

\clearpage
%\sout{{\bf (d)}, Progress of precision force measurement.  Both measurement sensitivity and the minimal measured force are shown for each existent work.  The $\blacktriangle$ and $\blacksquare$ symbols represent measurements based on microscopic particles~\cite{Biercuk2010,Shaniv2017,Blumseaao4453,Schreppler1486,Gilmore673} and macroscopic sensors~\cite{Moser2013,PhysRevLett.121.063602}, respectively. The black and green colors  mark alternating (a.c.) and static (d.c.) forces. The $\bDiamond$ symbol shows our result.

\begin{figure}
 \centerline{\includegraphics[angle=0,width= \textwidth]{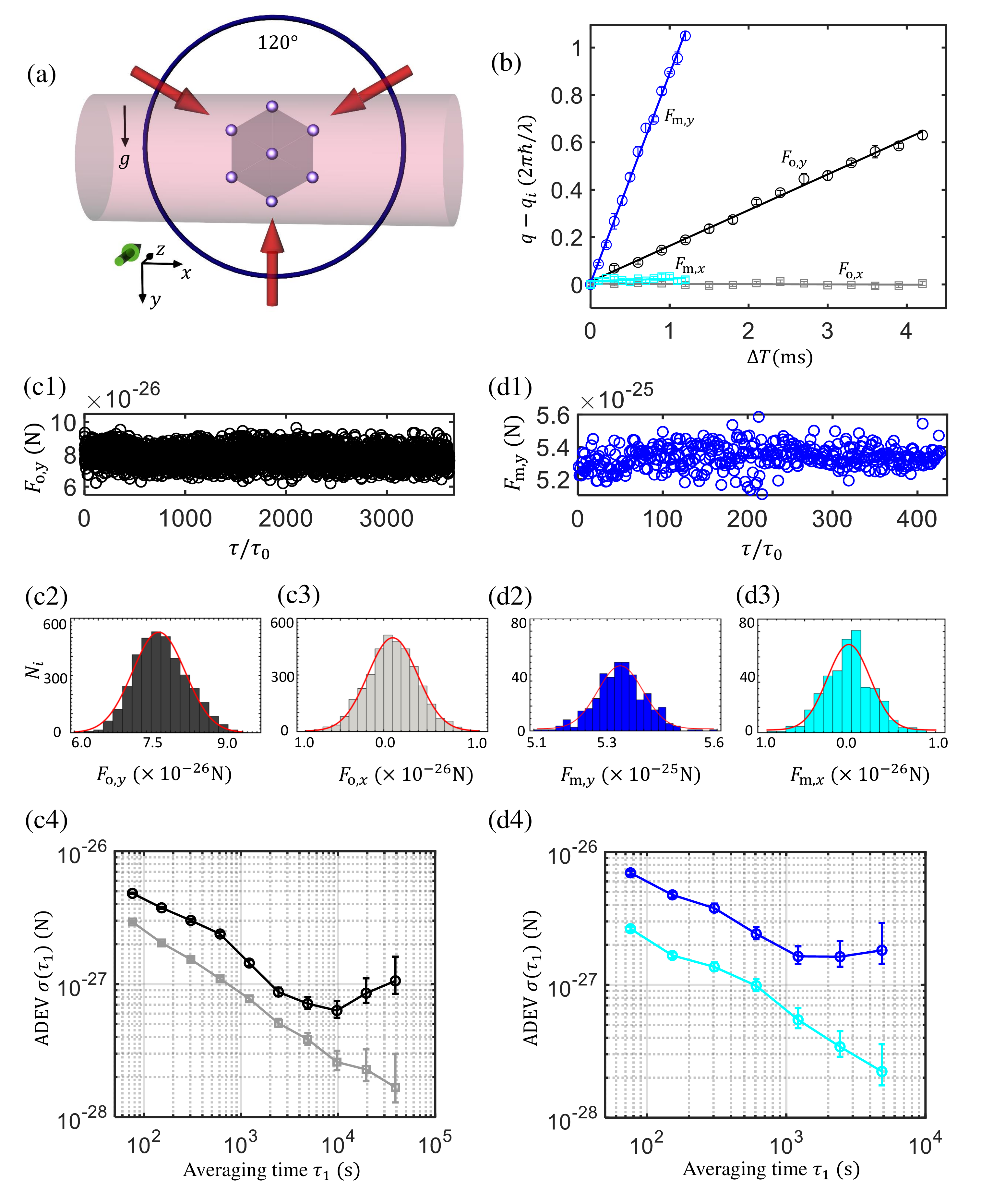}}
 \end{figure}
 \clearpage
\begin{figure}
\caption{ 	{\bf High-precision measurement of static electromagnetic  forces in two dimensions.} 
{\bf (a)}, The orientation of the atomic array confined by the triangular light crystal in the $x$-$y$ plane.  The system is probed by an imaging laser beam (green arrow) along the $\hat{z}$ direction. 
 {\bf (b)}, Measurement of atomic wave evolution upon static forces. Two types of forces, optical ($\vec{F}_{\rm o}$) and magnetic ($\vec{F}_{\rm m}$), have been implemented.  The wavevector of the BEC shows a linear dependence on the force acting time $\Delta T$.  The error bars represent the standard error averaging over five experimental runs. 
 {\bf (c)} and {\bf (d)} correspond to the experimental results on measuring the optical and magnetic forces, respectively.   Here $\tau$ is the total experimental time and $\tau_0 \approx 76$~s is  the time cost for a single data point. (c1) and (d1) present the raw data for the $y$-component forces.  (c2, c3, d2, d3) provide the histogram analysis of the measurements of the two spatial components of the forces.  (c4) and (d4) show the Allan deviation (ADEV) as a function of the averaging time ($\tau_1$), 
 with error bars representing $1\sigma$ statistical uncertainty. The grey and black (light blue and blue) colors illustrate $x$- and $y$-components of the optical  (magnetic) force.}
 		\label{fig:f2}
 	\end{figure}
\clearpage

\clearpage

 \clearpage 

%\end{singlespace} 

\end{document}